\documentstyle[epsfig,aps]{revtex}
\font\eightit=cmti8                     % Used in address list
\def\r#1{\ignorespaces $^{#1}$}         % Used in author and address list
\begin{document}
\draft
\onecolumn

%\wideabs{
\title{
\begin{center}
  Measurement of $d\sigma/dM$ and Forward-Backward Charge Asymmetry for 
  High Mass Drell-Yan $e^+e^-$ Pairs from $p\bar{p}$ Collisions at  
  $\sqrt{s}=1.8$~TeV
\end{center}
      }
\author{
%---CDF author list-----6/6/01-----------------------------------------------
%\font\eightit=cmti8
%\def\r#1{\ignorespaces $^{#1}$}
\hfilneg
\begin{sloppypar}
\noindent
T.~Affolder,\r {23} H.~Akimoto,\r {45}
A.~Akopian,\r {37} M.~G.~Albrow,\r {11} P.~Amaral,\r 8  
D.~Amidei,\r {25} K.~Anikeev,\r {24} J.~Antos,\r 1 
G.~Apollinari,\r {11} T.~Arisawa,\r {45}
% ...
\end{sloppypar}
\vskip .026in
\begin{center}
(CDF Collaboration)
\end{center}
}
\vskip .026in
\address{
\begin{center}
\r 1  {\eightit Institute of Physics, Academia Sinica, Taipei, Taiwan 11529, 
Republic of China} \\
\r 2  {\eightit Argonne National Laboratory, Argonne, Illinois 60439} \\
\r 3  {\eightit Istituto Nazionale di Fisica Nucleare, University of Bologna,
I-40127 Bologna, Italy} \\
\r 4  {\eightit Brandeis University, Waltham, Massachusetts 02254} \\
\r 5  {\eightit University of California at Davis, Davis, California  95616} \\
\r 6  {\eightit University of California at Los Angeles, Los 
Angeles, California  90024} \\  
\r 7  {\eightit Instituto de Fisica de Cantabria, CSIC-University of Cantabria, 
39005 Santander, Spain} \\
\r 8  {\eightit Enrico Fermi Institute, University of Chicago, Chicago, 
Illinois 60637} \\
\r 9  {\eightit Joint Institute for Nuclear Research, RU-141980 Dubna, Russia}
\\
\r {10} {\eightit Duke University, Durham, North Carolina  27708} \\
\r {11} {\eightit Fermi National Accelerator Laboratory, Batavia, Illinois 
60510} \\
\r {12} {\eightit University of Florida, Gainesville, Florida  32611} \\
\r {13} {\eightit Laboratori Nazionali di Frascati, Istituto Nazionale di Fisica
               Nucleare, I-00044 Frascati, Italy} \\
\r {14} {\eightit University of Geneva, CH-1211 Geneva 4, Switzerland} \\
\r {15} {\eightit Glasgow University, Glasgow G12 8QQ, United Kingdom}\\
\r {16} {\eightit Harvard University, Cambridge, Massachusetts 02138} \\
\r {17} {\eightit Hiroshima University, Higashi-Hiroshima 724, Japan} \\
\r {18} {\eightit University of Illinois, Urbana, Illinois 61801} \\
\r {19} {\eightit The Johns Hopkins University, Baltimore, Maryland 21218} \\
\r {20} {\eightit Institut f\"{u}r Experimentelle Kernphysik, 
Universit\"{a}t Karlsruhe, 76128 Karlsruhe, Germany} \\
\r {21} {\eightit Center for High Energy Physics: Kyungpook National
University, Taegu 702-701; Seoul National University, Seoul 151-742; and
SungKyunKwan University, Suwon 440-746; Korea} \\
\r {22} {\eightit High Energy Accelerator Research Organization (KEK), Tsukuba, 
Ibaraki 305, Japan} \\
\r {23} {\eightit Ernest Orlando Lawrence Berkeley National Laboratory, 
Berkeley, California 94720} \\
\r {24} {\eightit Massachusetts Institute of Technology, Cambridge,
Massachusetts  02139} \\   
\r {25} {\eightit University of Michigan, Ann Arbor, Michigan 48109} \\
\r {26} {\eightit Michigan State University, East Lansing, Michigan  48824} \\
\r {27} {\eightit University of New Mexico, Albuquerque, New Mexico 87131} \\
\r {28} {\eightit The Ohio State University, Columbus, Ohio  43210} \\
\r {29} {\eightit Osaka City University, Osaka 588, Japan} \\
\r {30} {\eightit University of Oxford, Oxford OX1 3RH, United Kingdom} \\
\r {31} {\eightit Universita di Padova, Istituto Nazionale di Fisica 
          Nucleare, Sezione di Padova, I-35131 Padova, Italy} \\
\r {32} {\eightit University of Pennsylvania, Philadelphia, 
        Pennsylvania 19104} \\   
\r {33} {\eightit Istituto Nazionale di Fisica Nucleare, University and Scuola
               Normale Superiore of Pisa, I-56100 Pisa, Italy} \\
\r {34} {\eightit University of Pittsburgh, Pittsburgh, Pennsylvania 15260} \\
\r {35} {\eightit Purdue University, West Lafayette, Indiana 47907} \\
\r {36} {\eightit University of Rochester, Rochester, New York 14627} \\
\r {37} {\eightit Rockefeller University, New York, New York 10021} \\
\r {38} {\eightit Rutgers University, Piscataway, New Jersey 08855} \\
\r {39} {\eightit Texas A\&M University, College Station, Texas 77843} \\
\r {40} {\eightit Texas Tech University, Lubbock, Texas 79409} \\
\r {41} {\eightit Institute of Particle Physics, University of Toronto, Toronto
M5S 1A7, Canada} \\
\r {42} {\eightit Istituto Nazionale di Fisica Nucleare, University of Trieste/
Udine, Italy} \\
\r {43} {\eightit University of Tsukuba, Tsukuba, Ibaraki 305, Japan} \\
\r {44} {\eightit Tufts University, Medford, Massachusetts 02155} \\
\r {45} {\eightit Waseda University, Tokyo 169, Japan} \\
\r {46} {\eightit University of Wisconsin, Madison, Wisconsin 53706} \\
\r {47} {\eightit Yale University, New Haven, Connecticut 06520} \\
\r {(\ast)} {\eightit Now at Carnegie Mellon University, Pittsburgh,
Pennsylvania  15213} \\
\r {(\ast\ast)} {\eightit Now at Northwestern University, Evanston, Illinois 
60208}
\end{center}
}
\maketitle
\wideabs{
\begin{abstract}
We report on a measurement of the mass dependence of the
forward-backward charge asymmetry, $A_{FB}$,
and production cross section $d\sigma/dM$ 
for $e^+e^-$ pairs with mass  $M_{ee}>40$~GeV$/c^2$. The data sample 
consists of 108~pb$^{-1}$ of $p\bar{p}$ collisions at $\sqrt{s}=1.8$~TeV
taken by the Collider Detector at Fermilab during 1992--1995.  
The measured asymmetry and  $d\sigma/dM$ are compared with the 
predictions of the  Standard Model and a model with an extra $Z'$ 
gauge boson. 
%The data are in good agreement with the Standard Model,
%but $A_{FB}$ at the highest mass values is
%2 $\sigma$ lower than the Standard Model prediction.
\end{abstract}
\pacs{PACS numbers: 13.85.Qk, 12.38.Qk, 12.15.Ji, 12.15.Mm, 12.60.-i,
12.60.Cn}
\twocolumn
}
%
%---Text-----------------------------------------------------------------------
%
%\par
%
In hadron-hadron collisions at high energies, massive $e^+e^-$ 
pairs are predominantly produced via the Drell-Yan~\cite{DrellYan} process. 
In the Standard Model, quark-antiquark annihilations form an 
intermediate $\gamma^*$ or $Z$ ($\gamma^*/Z$) vector boson, 
which may yield an $e^+e^-$ pair.
The presence of both vector
and axial-vector couplings in this process gives rise to a 
forward-backward asymmetry,
$A_{FB}$, in the final-state angle of the $\it {electron}$
in the rest frame of the $e^+e^-$ pair (with 
respect to the $\it{proton}$ direction). 
The Standard Model predicts accurately the mass [M] dependence of 
the production cross section ($d\sigma/dM$) and $A_{FB}$. 
In the region of the $Z$ resonance, the predicted
asymmetry is related to the electroweak mixing angle. For  $M \gg M_Z$, the 
large predicted asymmetry (near 0.6) is a consequence
of the interference between the propagators of the  $\gamma^*$ and $Z$. 
Various new interactions not included in the Standard Model could result
in deviations from the Standard Model predictions for 
both $d\sigma/dM$ and $A_{FB}$.  Possible new interactions include additional
gauge bosons~\cite{rosner,baur,zprime}, quark-lepton 
compositeness~\cite{ee}, exchange of R parity violating supersymmetric
particles or leptoquarks, and extra-dimensions~\cite{extra}.
\par
In this communication we extract measurements of 
$d\sigma/dM$ and improve the existing measurements of $A_{FB}$. 
The previous CDF measurements
in the   $e^+e^-$~\cite{ee}  and $\mu^+\mu^-$~\cite{DiMuPRD} channels
reported on a central $(|\eta|<1.1)$ region measurements of 
$d^{2}\sigma/{dMdy}$ averaged over rapidities~\cite{CdfDet} of $|y|<1$. 
Electrons in the plug $(1.1<|\eta|<2.4)$,
but not forward ($2.2<|\eta|<4.2$) calorimeter were included in the 
previous CDF measurement~\cite{afb} of $A_{FB}$, but backgrounds in the plug 
region were large resulting in significant uncertainties. 
This analysis includes electrons in the plug
and forward regions of the calorimeter, thus covering
a larger range of rapidity.
 New analysis techniques and 
tracking requirements in the
forward region greatly reduce
the backgrounds and uncertainties.
In addition, by applying electroweak and  QED radiative
corrections~\cite{QEDRadCor,Photos20} (which correct for
the change in M from the emission of final state photons), measurements 
of $A_{FB}$ in small bins over a large range in mass are extracted
for the first time. Our measurement of $e^+e^-$ pairs with 
%for a range quark-antiquark 
invariant masses between 40 and 500 GeV$/c^2$ complement measurements 
of cross-sections and asymmetries in $e^+e^-$ machines, which currently 
extend to 200 GeV$/c^2$.
\par
The $e^+e^-$ pairs come from 108~pb$^{-1}$ of $p\bar{p}$ collisions
at $\sqrt{s}=1.8$~TeV taken by the Collider Detector at Fermilab~\cite{CdfDet}
(CDF) during 1992--1993 ($18.7 \pm 0.7$~pb$^{-1}$) and 1994--1995 
($89.1 \pm 3.7$~pb$^{-1}$). The CDF detector consists of a 
solenoidal magnetic spectrometer 
surrounded by projective-tower-geometry calorimeters and outer muon 
detectors. Only detector components used in this measurement are described 
here. Charged particle momenta and directions are measured by the 
spectrometer embedded  in a 1.4~T axial magnetic field, an 84-layer
cylindrical drift chamber (CTC), an inner vertex tracking chamber (VTX),
and a silicon vertex detector (SVX). The polar coverage of the CTC 
tracking is $|\eta|<1.2$. The $p\bar{p}$ collision point along the beam 
line $(Z_{\rm vertex})$ is determined using tracks in the VTX. 
The energies and directions~\cite{CdfDet} of electrons, photons, 
and jets are measured by three separate calorimeters covering three regions: 
central, end plug, and forward. Each calorimeter has an 
electromagnetic (EM) and a hadronic (HAD) calorimeter. 
%
%\par
%
We follow the analysis procedure used in our previous 
measurements~\cite{liu,willis} of $d\sigma/dy$ and $d\sigma/dP_{\rm T}$ of $Z$
boson pairs. 
The measurement of $A_{FB}$ requires the sign of at least one lepton.
Therefore, we require one of the leptons to be in the central region.
\par
The sample of $e^+e^-$ events was collected by a three-level online 
trigger that required an electron in either the central or the plug 
calorimeter. The offline analysis selected events with two or more 
electron candidates.
Since the electrons from the Drell-Yan process are typically isolated,
a cut on the amount of transverse energy in
a cone outside the electron shower is imposed.
Electrons in the central, end plug, and forward regions 
are required to be within the fiducial area
of the calorimeters and to have a minimum
$E_{\rm T}$ of 22, 20, and 15 GeV, respectively. 
To improve the purity of the sample, electron identification cuts
are applied~\cite{EMuSel}. 
%For CC, CP, and CF events, 
The central electron (or one of them if 
there are two) is required to pass strict criteria. The criteria 
on the other electron are looser. A central electron must have a 
CTC track that extrapolates to the electron's shower cluster in 
the EM calorimeter. This cluster must have EM-like transverse 
shower profiles. The track momentum and the EM shower energy must 
be consistent with one another. The track is also used to determine 
the position and direction of the central electron. The fraction of 
energy in the HAD calorimeter towers behind the EM shower is required 
to be consistent with that expected for an EM shower 
($E_{\rm HAD}/E_{\rm EM}$). 
The plug electrons must also have an EM-like transverse 
shower profile. The end plug and forward electrons 
are required to pass the $E_{\rm HAD}/E_{\rm EM}$ requirement and to
have a track in the VTX which originates from the same vertex as the 
other electron and points to the position of the electromagnetic cluster 
in the calorimeter. The ratio of found to expected hits in the VTX
is required to be greater than $70\%$ and $50\%$ for plug and 
forward electrons, respectively. The VTX tracking efficiency is 
$(97.8\pm 0.3)\%$ for plug electrons and $(97.0\pm 0.9)\%$ for 
forward electrons. 
To further reduce the backgrounds
from dijet events for very forward electrons 
both leptons legs for CF
events are required to be on the same side of the detector
i.e. have the same sign in $\eta$.
The total data sample after all cuts consists of 7632 events.
\par
With the above cuts, backgrounds are low and can 
be reliably estimated using the data. Because of the CTC tracking 
requirement, the jet background for the CC sample is negligible.
All CC events are required to be opposite sign (the 8 same sign events
are all in the $Z$ mass region and originate from the small charge 
misidentification probability).
The CC background is mainly composed of $e^+e^-$ pairs from $W^+W^-$, 
$\tau^+\tau^-$, $c\bar{c}$, $b\bar{b}$, and $t\bar{t}$ sources. 
This background is estimated using $e^\pm\mu^\mp$ pairs. 
 %~\cite{EMuSel} (the highest $e^\pm\mu^\mp$ pair mass is 118 GeV$/c^2$). 
Because of the tracking and same side requirements,
the backgrounds in the CP and CF
% PP, and PF
topologies are also small and mainly come from
dijets events. These are extracted from the data by fitting the isolation
energy distribution to a sum of the expected distributions
for signal and background events. For $M_{ee}>105$ GeV the 
isolation energy fraction is used. The shapes of the isolation energy
distributions are extracted from  specially
selected jet background-dominated and electron signal-dominated data samples. 
\par
The acceptance for Drell-Yan
$e^+e^-$ pairs is obtained using the Monte Carlo event generator
{\small PYTHIA 6.146}~\cite{Pythia}, and CDF detector simulation programs.
{\small PYTHIA} generates the LO QCD
interaction ($q+\bar{q} \rightarrow \gamma^*/Z$), simulates initial state QCD
radiation via its parton shower algorithms, and generates the decay,
$\gamma^*/Z \rightarrow e^+e^-$. To approximate higher order QCD corrections
to the LO mass distribution, a $K$-factor~\cite{AEM-Kfactor}
is used as an event weight:
$K(M^2) = 1 + \frac{4}{3}(1 + \frac{4}{3}\pi^2) \alpha_s(M^2)/2\pi$,
where $\alpha_s$ is the QCD coupling. This factor improves the
agreement between the NLO and LO mass spectra.
(For $M>50$ GeV/$c^2$, $1.25 < K < 1.36$.)
The CTEQ5L~\cite{CTEQ5} PDFs are used in the acceptance calculations. 
Final state QED 
radiation~\cite{QEDRadCor} from the $\gamma^*/Z \rightarrow e^+e^-$ vertex is
added by the {\small PHOTOS}~\cite{Photos20} Monte Carlo program.
The parameters of {\small PYTHIA 6.146} are tuned~\cite{Pythia}
to fit experimentally measured transverse momentum~\cite{willis} and rapidity
distributions~\cite{liu} of $e^+e^-$ pairs in the $Z$ region.
Generated events are
processed by CDF detector simulation programs and are reconstructed as data.
The calorimetry energy scales, resolutions,
cut efficiencies and
corresponding errors used in the detector
simulation are extracted from the data.
Reconstructed Monte Carlo events are accepted if
they pass the fiducial, kinematic, and mass cuts.
\par
\begin{figure}
\begin{center}
\mbox{\epsfxsize=3.37in \epsfysize=4.00in \epsffile{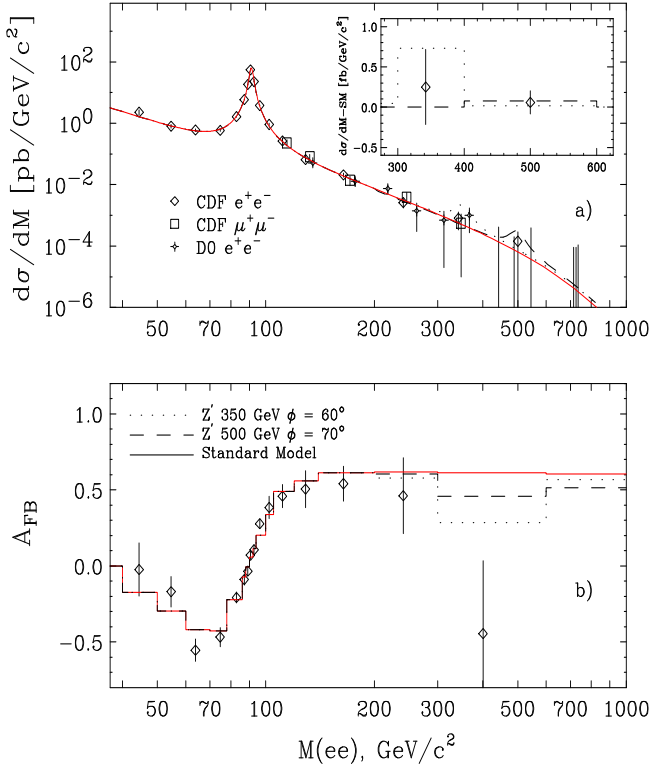}}
\end{center}
\caption{(a) $d\sigma/dM$ 
distribution of $e^+e^-$  and $\mu^+\mu^-$ pairs. All errors 
(except for the overall 3.9 $\%$ luminosity error)
have been combined in quadrature.The Standard Model
theoretical predictions (solid line) have been normalized by a factor
of 1.11 to the data in the
$Z$ boson mass region. Also shown are  $e^+e^-$ measurements
from \mbox{DO\hspace*{-1.5ex}/}. 
(b)  $A_{FB}$ versus mass compared to the 
 Standard Model expectation (solid line).
Also, predicted theoretical curves for $d\sigma/dM$ and $A_{FB}$
with an extra $E_6$ $Z'$ boson (width of 10$\%$)
with $M_{Z'}$=350 (dotted line) and 500 GeV$/c^2$
(dashed line).
%for $\phi$=$70^0$ (solid) and $\phi$=$160^0$ (dotted).
The inset in (a) shows the  difference between the
$d\sigma/dM$ CDF data for $e^+e^-$ pairs and the Standard Model prediction (on
a linear scale) compared to expectation from these two $Z'$ models.
}
\label{fig1}
\end{figure}

The angle of the electron is determined in the 
Collins-Soper~\cite{collins} frame, which reduces the uncertainty in the 
angle introduced by the transverse momentum of the incoming partons. 
We determine the cross sections $d\sigma^+/dM$
and $d\sigma^-/dM$ for forward and backward 
events, respectively. $d\sigma/dM$(shown in Figure~\ref{fig1}(a)) and 
$A_{FB}$ (shown in Figure~\ref{fig1}(b)) are extracted from the
sum and difference of the forward and backward cross sections.
The very small misidentification probability is handled as
a dilution factor in $A_{FB}$.
%In the $d\sigma/dM$ measurement (shown in Figure~\ref{fig1}(a)),
The CC, CP, and CF
%, PP and PF
samples are  
combined and are binned in $M$, and $d\sigma/dM$ is given by
\begin{displaymath}
 \frac{d\sigma}{dM} =
 \frac{N-B} {C_{rc}\Delta M  \:
                            \sum_r {\cal L}_r \: \epsilon A_r} .
\end{displaymath}
Here, $N$ is the number of observed events, $B$ is the estimated
background in a bin, $C_{rc}$ is the radiative 
correction~\cite{QEDRadCor,Photos20} factor
to convert the measured cross sections to a Born level (one boson exchange)
cross section, 
$\Delta M$ is the bin width, the sum
$r$ is over the 1992--1993 and 1994--1995 runs, ${\cal L}_r$ is the 
integrated luminosity, and $\epsilon A_r$ is the run's combined event selection
efficiency and acceptance.
The backgrounds subtracted from the event count are
predicted using the data and background samples. 
Acceptances are calculated separately for CC, CP, and CF
%, PP, and PF
pairs. They are combined with the corresponding event selection efficiencies 
to give $\epsilon A_r$. $A_r$ includes 
corrections for detector resolution smearing.
Since the {\small PYTHIA/PHOTOS}~\cite{Photos20} Monte 
Carlo includes radiative final state photons, 
%using a colinear approximation,
the effects of the final state radiated photons are 
calculated together with the acceptance calculation.
Also shown in Figure~\ref{fig1}(a) are CDF dimuon results for $d\sigma/dM$  
(which have not been previously published). These were extracted 
from the published dimuon~\cite{DiMuPRD}
data ($d^{2}\sigma/{dMdy}$ for $|y|<1$) by applying a correction
using a QCD-NLO calculation~\cite{NLO} that describes 
our measured $Z$ $e^+e^-$ pair
$y$ distributions~\cite{liu} over the entire kinematic range.
The $e^+e^-$ cross sections and asymmetries for $M_{ee}<105$ GeV$/c^2$ are 
given in Table~\ref{Table2}, and the higher mass data samples and backgrounds 
are summarized in Table~\ref{Table1}. 
We define $\overline{M}$ to be the mass value for which the
cross section is equal to the average cross section over each bin, as
calculated from a NNLO theoretical prediction~\cite{NNLO} for $d\sigma/dM$.
\par
The systematic errors are determined from the following: statistical
errors in the fits to the fractional background, variations in the background
estimates using different methods, the background in the efficiency
sample, the uncertainty in energy resolution of the calorimeter,   
the choice of PDFs, and the distribution of $e^+e^-$ $P_{\rm T}$ used 
in the Monte Carlo event generator.  Above the region of the $Z$ mass, the 
total systematic errors are much smaller than the statistical errors,
and the total errors shown in Table~\ref{Table1} are dominated by statistics. 
In the $Z$ mass region, the systematic errors are less than half of the
statistical error (as shown below).
The $p\bar{p}$ collision luminosity is derived with CDF's beam-beam
cross section, $\sigma_{\rm BBC}=51.15 \pm 1.60$~mb~\cite{CDFLum,CDFLum1}.
The luminosity error of 3.9$\%$ contains the $\sigma_{\rm BBC}$ 
error and uncertainties specific to running conditions.
As described in a previous communication~\cite{liu},
the extracted cross section in the $Z$ mass region is 
$252.1 \pm 3.9\:{\rm (stat.)} \pm 1.6\:{\rm (syst.)} \pm 9.8\:{\rm (lum.)}$~pb.
Note that since the $p\bar{p}$ inelastic cross section used by CDF in 
luminosity calculations differs 
from \mbox{DO\hspace*{-1.5ex}/}{\hspace*{0.5ex}'s} by 
$+5.9\%$~\cite{CDFLum1}, the \mbox{DO\hspace*{-1.5ex}/} 
$e^+e^-$ cross sections~\cite{D0} shown in Figure~\ref{fig1}
have been multiplied by 1.059.
\par
Figure~\ref{fig1} also compares the measured 
$d\sigma/dM$ and $A_{FB}$ to theoretical predictions.
The $d\sigma/dM$ curve is a QCD NNLO~\cite{NNLO} calculation
with MRST99 NLO PDFs~\cite{MRS}. 
The predictions in Figure~\ref{fig1}(a) are normalized by a 
factor F=1.11, the ratio of measured total cross section in the $Z$ region
to  the NNLO prediction (the overall normalization uncertainties 
are 3.9$\%$ for the experimental data and $5\%$ for the NNLO theory).
The Standard Model predictions for $A_{FB}$ have been
calculated~\cite{baur,QEDRadCor} in NLO-QCD.
%with $sine^2\theta_{eff}=0.232$.
%
%
The measured 
$d\sigma/dM$ and $A_{FB}$ values are in good agreement with the Standard Model 
predictions.  However, $A_{FB}$ in the highest
mass bin (300-600 GeV$/c^2$) is 2.2$\sigma$ (standard
deviations)  below the Standard Model prediction.  There are
three events~\cite{events}
in the negative hemisphere and one event in the positive
hemisphere.
A negative asymmetry in this region
could result from  new interactions including additional
gauge bosons~\cite{rosner,baur,zprime} (discussed as an example below),
quark-lepton compositeness~\cite{ee},
exchange of R parity violating supersymmetric
particles, leptoquarks, and extra-dimensions~\cite{extra}. 
For example,
the predicted theoretical curves~\cite{baur} for $d\sigma/dM$ and $A_{FB}$
in models which include
extra E6 $Z'$ bosons (with parameters tuned to fit low energy 
electroweak data)
are shown as the dashed and dotted curves in Figure~\ref{fig1}. 
\par
In summary, the measured 
$d\sigma/dM$ and $A_{FB}$ values are in good agreement with the Standard Model 
predictions. At the highest mass values, the measured value of $A_{FB}$ is
2 $\sigma$
lower than the Standard Model prediction. These high mass Drell-Yan data
will be included in global fits to electroweak data to search (or
extract limits) for physics beyond the Standard Model.
\par
We thank the Fermilab staff and the technical staffs of the
participating institutions for their vital contributions.  This work was
supported by the U.S. Department of Energy and National Science Foundation;
the Italian Istituto Nazionale di Fisica Nucleare; the Ministry of Education,
Science, Sports and Culture of Japan; the Natural Sciences and Engineering 
Research Council of Canada; the National Science Council of the Republic of 
China; the Swiss National Science Foundation; the A. P. Sloan Foundation; the
Bundesministerium fuer Bildung und Forschung, Germany; the Korea Science 
and Engineering Foundation (KoSEF); the Korea Research Foundation; and the 
Comision Interministerial de Ciencia y Tecnologia, Spain.
%Additional details of this analysis are given in CDF note 5475 and
%CDF note 5471.
%-----------------------------------------------------------------------------

%
\begin{table}
\caption{The $e^+e^-$ cross sections and asymmetries 
for $M_{ee}<105$~GeV$/c^2$. The statistical and systematic errors 
have been combined in quadrature. The $3.9\%$ overall luminosity 
error is not included.}
\label{Table2}
\begin{center}
\begin{tabular}{ccrr}
Mass Bin & $\overline{M_{ee}}$ & 
\multicolumn{1}{c}{$d\sigma_{ee}/dM$} & \multicolumn{1}{c}{$A_{FB}$} \\
(GeV$/c^2$)& (GeV$/c^2$)  &(pb/GeV$/c^2$)  &  \\ \hline
   $40-50$ &  44.5  & $  2.30 \pm  0.47 $  & $ -0.02 \pm  0.17 $  \\
   $50-60$ &  54.6  & $  0.80 \pm  0.11 $  & $ -0.17 \pm  0.10 $  \\
   $60-70$ &  63.8  & $  0.60 \pm  0.06 $  & $ -0.56 \pm  0.07 $  \\
   $70-78$ &  74.7  & $  0.58 \pm  0.05 $  & $ -0.47 \pm  0.06 $  \\
   $78-86$ &  82.9  & $  1.64 \pm  0.06 $  & $ -0.21 \pm  0.04 $  \\
   $86-88$ &  87.1  & $  5.89 \pm  0.22 $  & $ -0.09 \pm  0.04 $  \\
   $88-90$ &  89.2  & $ 18.28 \pm  0.52 $  & $ -0.04 \pm  0.03 $  \\
   $90-92$ &  90.6  & $ 54.85 \pm  1.39 $  & $  0.07 \pm  0.02 $  \\
   $92-94$ &  92.8  & $ 22.75 \pm  0.66 $  & $  0.11 \pm  0.03 $  \\
  $94-100$ &  96.2  & $  3.80 \pm  0.12 $  & $  0.28 \pm  0.03 $  \\
 $100-105$ & 102.2  & $  0.91 \pm  0.07 $  & $  0.38 \pm  0.07 $  \\ 
\end{tabular}
\end{center}
\end{table}
\begin{table}
\caption{The number of signal and background events
and correction factors involving the extraction
%of $d\sigma_{ee}^+/dM$ (forward hemisphere), $d\sigma_{ee}^-/dM$
%(backward hemispher),
of $d\sigma_{ee}^+/dM$, $d\sigma_{ee}^-/dM$,
$d\sigma_{ee}/dM$,
and $d\sigma_{\mu\mu}/dM$ for mass
bins above the $Z$ pole.
%The mass bins are defined in the text
%with $\overline{M}$,
% is the theoretically expected average mass (in GeV$/c^2$)
%for forward, backward and all events, respectivel.
 Here, $N$ is the number of events, $B$ is the background estimate, 
$\epsilon AC_{rc}$ is the combined efficiency, acceptance,  
smearing and radiative correction. The statistical and systematic
errors for $d\sigma/dM$ have been combined in quadrature.
The $3.9\%$ overall luminosity error is not included.}
\label{Table1}
\begin{center}
\begin{tabular}{ccrrcr}
\multicolumn{1}{c}{Mass Bin} &
\multicolumn{1}{c}{$\overline{M_{ee}}$} & \multicolumn{1}{c}{$N$} &
\multicolumn{1}{c}{$B$} & \multicolumn{1}{c}{$\epsilon AC_{rc}$} &
   \multicolumn{1}{c}{$d\sigma_{ee}^+/dM$} \\ 
(GeV$/c^2$) & (GeV$/c^2$) & & & & (fb/GeV$/c^2$) \\ \hline
   $105-120$ & 111.2  & 93 &  2.3 & 0.293 & $  191.55\pm 20.38 $  \\
   $120-140$ & 128.8  & 34 &  1.4 & 0.307 & $   49.20\pm  8.81 $  \\
   $140-200$ & 164.2  & 35 &  1.2 & 0.326 & $   16.01\pm  2.80 $  \\
   $200-300$ & 240.6  &  7 &  0.1 & 0.337 & $    1.88\pm  0.73 $  \\
   $300-400$ & 343.2  &  1 &  0.0 & 0.330 & $    0.28\pm  0.28 $  \\
   $400-600$ & 478.8  &  0 &  0.0 & 0.307 & $    0.00\pm  0.15 $  \\
   $600-999$ & 725.6  &  0 &  0.0 & 0.227 & $    0.00\pm  0.10 $  \\\hline
\multicolumn{1}{c}{Mass Bin} &
\multicolumn{1}{c}{$\overline{M_{ee}}$} & \multicolumn{1}{c}{$N$} &
\multicolumn{1}{c}{$B$} & \multicolumn{1}{c}{$\epsilon AC_{rc}$} &
   \multicolumn{1}{c}{$d\sigma_{ee}^-/dM$} \\ \hline
   $105-120$ & 111.2  & 39 &  2.3 & 0.321 & $   70.64\pm 12.04$  \\
   $120-140$ & 128.8  & 13 &  1.4 & 0.334 & $   16.07\pm  5.01$  \\
   $140-200$ & 164.2  & 12 &  1.2 & 0.353 & $    4.74\pm  1.52$  \\
   $200-300$ & 240.6  &  3 &  0.1 & 0.386 & $    0.69\pm  0.42$  \\
   $300-400$ & 343.2  &  2 &  0.0 & 0.398 & $    0.47\pm  0.33$  \\
   $400-600$ & 478.8  &  1 &  0.0 & 0.391 & $    0.12\pm  0.12$  \\
   $600-999$ & 725.6  &  0 &  0.0 & 0.355 & $    0.00\pm  0.07$  \\\hline
\multicolumn{1}{c}{Mass Bin} &
\multicolumn{1}{c}{$\overline{M_{ee}}$} & \multicolumn{1}{c}{$N$} &
\multicolumn{1}{c}{$B$} & \multicolumn{1}{c}{$\epsilon AC_{rc}$} &
 \multicolumn{1}{c}{$d\sigma_{ee}/dM$} \\ \hline
  $105-120$ & 111.2  &132 &  4.7 & 0.300 & $  262.29\pm 23.67$  \\
  $120-140$ & 128.8  & 47 &  2.9 & 0.313 & $   65.46\pm 10.17$  \\
  $140-200$ & 164.2  & 47 &  2.4 & 0.331 & $   20.81\pm  3.20$  \\
  $200-300$ & 240.6  & 10 &  0.3 & 0.346 & $    2.60\pm  0.85$  \\
  $300-400$ & 342.2  &  3 &  0.0 & 0.343 & $    0.81\pm  0.47$  \\
  $400-600$ & 478.8  &  1 &  0.0 & 0.324 & $    0.14\pm  0.14$  \\
  $600-999$ & 725.6  &  0 &  0.0 & 0.252 & $    0.00\pm  0.09$  \\ \hline
\multicolumn{1}{c}{Mass Bin} &
\multicolumn{1}{c}{$\overline{M_{\mu\mu}}$} & \multicolumn{1}{c}{$N$} &
\multicolumn{1}{c}{$B$} & \multicolumn{1}{c}{$\epsilon AC_{rc}$} &
   \multicolumn{1}{c}{$d\sigma_{\mu\mu}/dM$} \\ \hline
   $110-120$ & 114.5  & 29 &  0 & 0.181 &  $ 231.93 \pm 51.56$  \\
   $120-150$ & 132.5  & 28 &  0 & 0.167 &  $  80.61 \pm 15.93$  \\
   $150-200$ & 171.2  &  9 &  0 & 0.164 &  $  13.98 \pm  4.67$  \\
   $200-300$ & 240.6  &  6 &  0 & 0.168 &  $   3.91 \pm  1.58$  \\
   $300-400$ & 343.2  &  1 &  0 & 0.176 &  $   0.56 \pm  0.56$  \\
   $400-600$ & 478.8  &  0 &  0 & 0.227 &  $   0.00 \pm  0.20$  \\
   $600-999$ & 725.6  &  0 &  0 & 0.230 &  $   0.00 \pm  0.11$  \\ 
\end{tabular}
\end{center}
\end{table}

\begin{references}
\bibitem{events} The (mass,$cos\theta$,region) of the four highest
mass $ee$ events are (303,0.4,CP), (319,$-$0.41,CC), (354,$-$0.14,CC),
and (496,$-$0.29,CC),  respectively.
 Because of acceptance of the muon chambers is limited to
  the central region, a good measurement of $A_{FB}$ cannot be
   done in this channel. Nonetheless, the three highest mass dimuon
   events (292, 300, and 320 GeV) all have positive $cos\theta$.
\end{references}
\end{document}